\documentclass[letterpaper,10pt]{article}
\usepackage{osameet3}

\usepackage{amsmath,amssymb}

\begin{document}

\title{Apparatus to Measure Optical Scatter of Coatings Versus Annealing Temperature}

\author{Joshua R. Smith$^1$, Rana X Adhikari$^2$, Katerin M. Aleman$^1$, Adrian Avila-Alvarez$^1$, Garilynn Billingsley$^2$, Amy Gleckl$^1$, Jazlyn Guerrero$^1$, Ashot Markosyan$^4$, Steven Penn$^3$, Juan A. Rocha$^1$, Dakota Rose$^1$, Robert Wright$^1$}
\address{$^1$Gravitational-Wave Physics and Astronomy Center, California State University Fullerton, Fullerton, CA 92831, USA}
\email{josmith@fullerton.edu}
\address{$^2$LIGO, California Institute of Technology, Pasadena, California 91125, USA}
\address{$^3$Hobart and William Smith Colleges, Geneva, New York 14456, USA}\address{$^4$Stanford University, Stanford, California 94305, USA}

\begin{abstract}
Light scattered by amorphous thin-film optical coatings limits the sensitivity of interferometric gravitational-wave detectors. We describe an imaging scatterometer to assess the role that crystal growth during annealing plays in this scatter. 
\end{abstract}


%\begin{document}

\maketitle

\section{Introduction}\label{sec:intro}
The Advanced Laser Interferometer Gravitational-Wave Observatory (aLIGO) has opened a new window on the universe by observing gravitational waves from merging black holes and neutron stars~\cite{gw150914,gw170817}. 
The aLIGO observatories in Hanford, WA, and Livingston, LA, are each dual-recycled Michelson interferometers with 4\,km-long Fabry-Perot arm cavities~\cite{aligo}. Light scattering from the optics of the aLIGO detectors (and their international partners Advanced Virgo~\cite{avirgo} and KAGRA~\cite{kagra}) increases the optical loss in the cavities, reducing the power buildup and thus degrading the detectors' shot-noise-limited sensitivity~\cite{aligo}. Additionally, stray light bounces off of moving objects such as vacuum chamber walls and re-enters the interferometer with arbitrary phase, adding noise~\cite{virgoscatter}. 
% adding nonlinear noise 
% Additionally, in the near future the efficacy of using squeezed states of light will be limited by scattering losses in the optics used to prepare and inject the squeezed light. 

aLIGO's core optics are superpolished fused silica substrates ion-beam sputtered with alternating near-quarter-wavelength layers of amorphous silica, SiO$_2$, and titania-doped tantala, Ti:Ta$_2$O$_5$, (low and high index, respectively) with the number and thickness of the layers chosen to give the optics their desired optical properties at 1064\,nm while keeping the coating's mechanical loss and thus thermal noise low~\cite{aligo,Abernathy:14,Harry:2011book}. After coating, these optics are annealed in air with a ramp of roughly 1$^{\circ}$C/minute to $500^{\circ}$C and held at that temperature (soaked) for 24 hours to improve their optical absorption and mechanical properties.

The scattering of these optics predicted by Rayleigh-Rice perturbation theory~\cite{Stover:2012book} based on their surface roughness is generally less than a few ppm (averaged surface roughness of $\sigma=$0.13\,nm RMS~\cite{coreoptics} predicts %$TIS=P_{scattered}/P_{specular}
%\approx(4\pi\cos{\theta_{incidence}}\sigma/\lambda)^2=
total scatter of $2.4$\,ppm), while measured scatter losses of the optics are significantly higher (measured average is 9.5\,ppm~\cite{coreoptics}). Additionally, images of scattering from the surfaces of these optics show constellations of point-like scattering. The origin of this larger than expected scatter is unknown. One plausible cause is micro-crystals in the otherwise amorphous Ti:Ta$_2$O$_5$ that nucleate or grow from existing nucleation sites~\cite{MARSEGLIA198031} during the preparation of the optics.

The onset of crystal nucleation and growth in amorphous Ta$_2$O$_5$ thin films is well studied, but varied. Un-doped Ta$_2$O$_5$ films crystallize between 600$^{\circ}$C and 850$^{\circ}$C~\cite{atanassova2003high}, while Ti:Ta$_2$O$_5$ shows no signs of crystallization at 600$^{\circ}$C (see \cite{Abernathy:14} and references therein). 
%However, already at 600$^{\circ}$C for 30-minute oxygen anneals, electron diffraction analysis indicates 
% (001) 
%crystal nucleation in Ta$_2$O$_5$. 
It is not known whether the LIGO annealing process triggers crystal growth.

To test whether annealing increases scatter from LIGO optics, we have developed an experiment to illuminate and image coated optics while they are being annealed and search for changing characteristics (such as new or larger point scatterers) as a function of temperature.

\section{Experiment design}
The purpose of the experiment is to assess scatter versus annealing temperature by mounting a coated sample optic in vacuum and annealing it to temperatures up to 600$^{\circ}$C, while illuminating it with a light source similar to LIGO's (1064\,nm) and recording images of the scatter from its coating. 

\begin{figure}[ht]
\begin{minipage}[t]{0.49\textwidth}
	\includegraphics[width=1.0\textwidth]{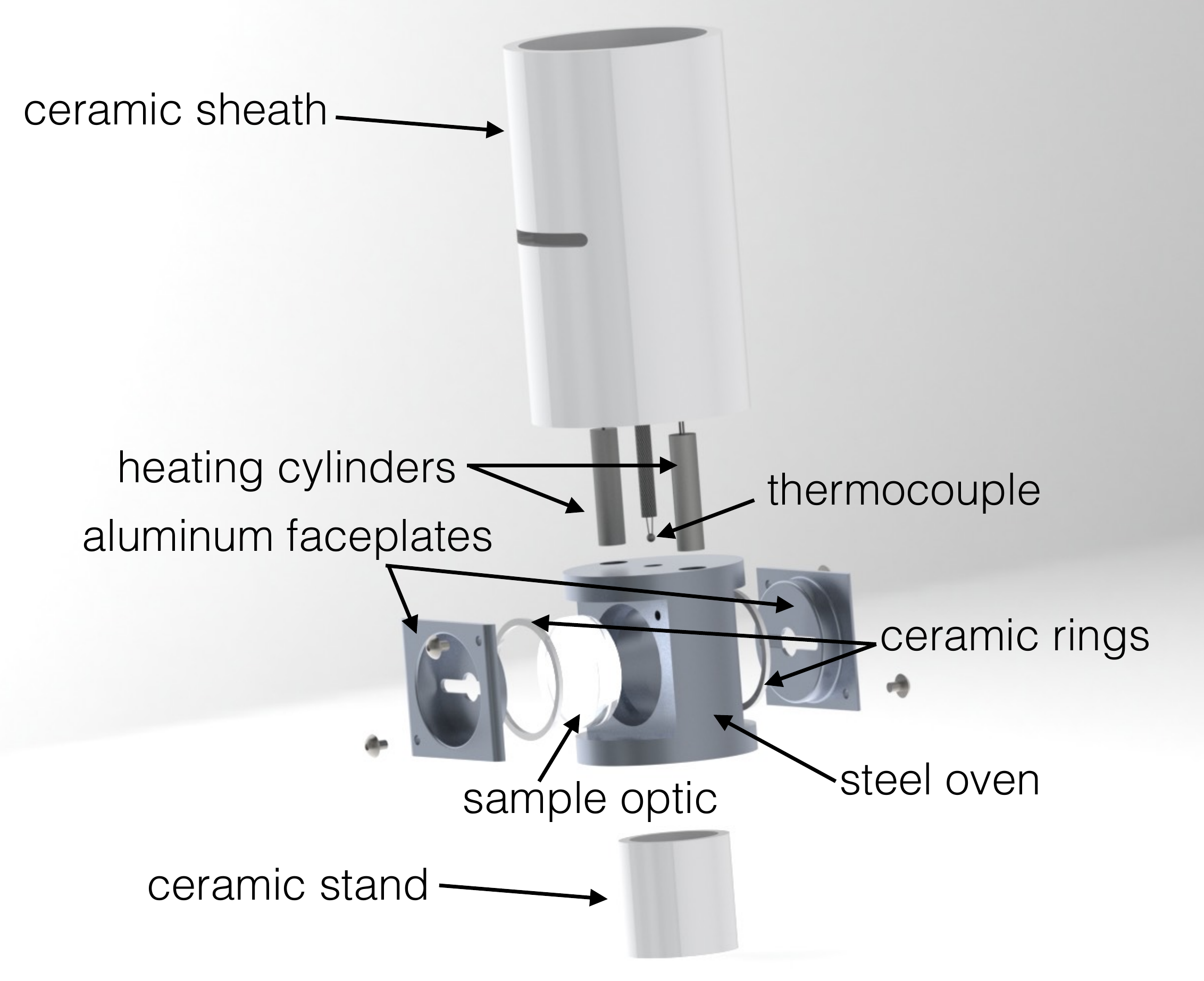}
\end{minipage}
\begin{minipage}[t]{0.51\textwidth}
	\includegraphics[width=1.0\textwidth]{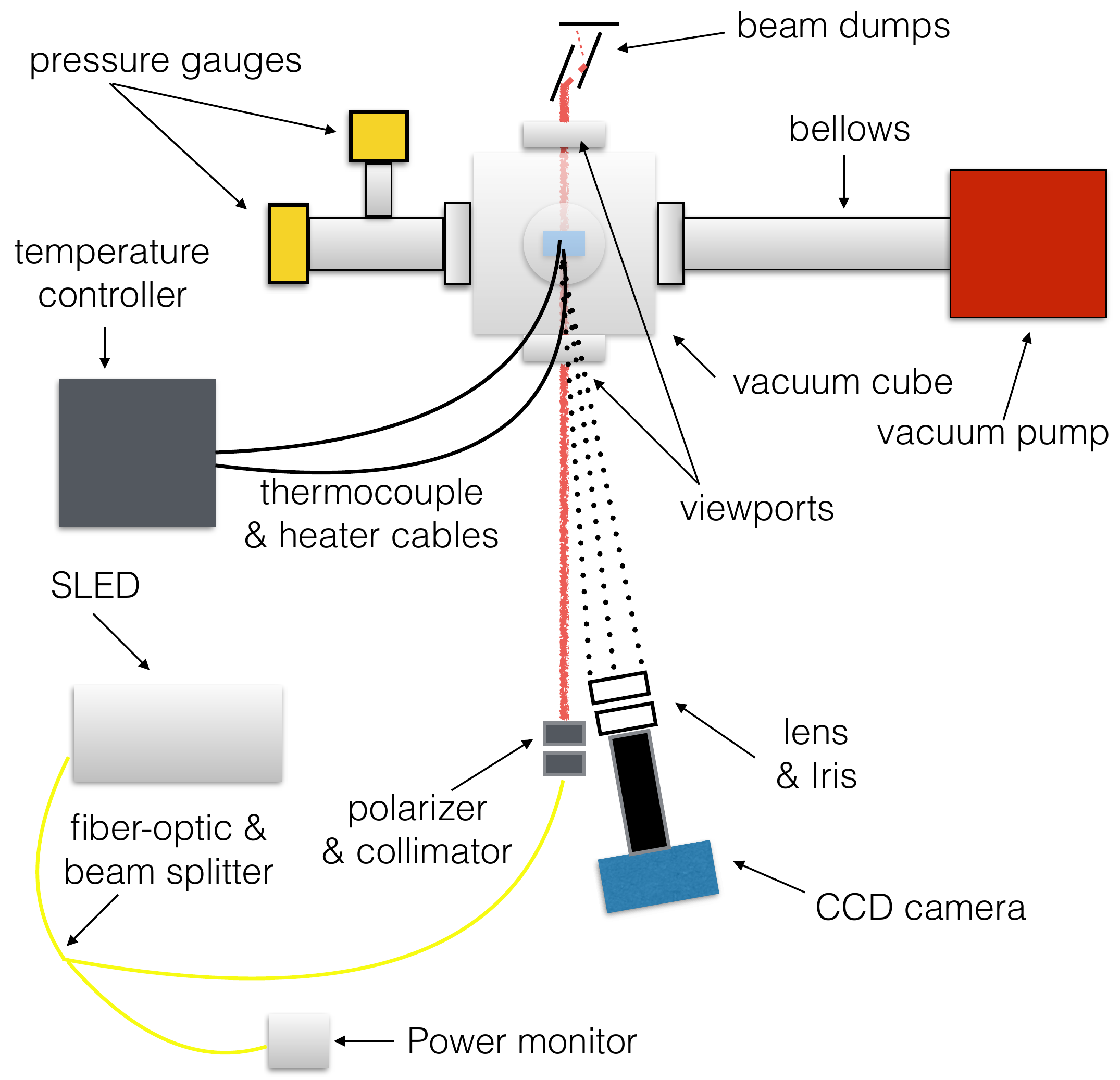}
\end{minipage}
\caption{Left: Exploded view of apparatus to hold and heat the sample. Right: Overhead diagram of the full setup including the vacuum chamber and the laser and imaging optics.}\label{fig:oven}
\end{figure}

The setup is shown in Figure~\ref{fig:oven}. The entire experiment takes place on a 6'x4' optical bench with a laminar-flow softwall cleanroom to reduce dust. To reduce coherent-light effects such as twinkling and speckle, a broader line-width 1050\,nm Super-Luminescent Diode (SLED) is used as the light source, at near normal incidence. A 1-inch optic is mounted in a steel oven with thin ceramic rings near its edges and covered front and back with aluminum faceplates that each have a rectangular opening with 5\,mm diameter circular apertures at their center for the incident and transmitted laser beam to shine though and for the camera to image the optic. The oven is heated by two cylindrical heaters, driven by a high-voltage feedback controller, and the optic's temperature is measured using a thermocouple. The oven sits on a ceramic stand, and is surrounded by a ceramic sheath, to decouple it thermally from the steel vacuum cube. The experiment is conducted under high vacuum. 

Light scattered from the optic's surface is imaged at 10$^{\circ}$ from  incidence using a single converging lens and iris to cast an image onto a low-noise 4096x4096 CCD camera, as described in~\cite{vander-hyde2015,Magana-Sandoval:12}. Images are taken once every 300\,s while the sample is annealed using ramp and soak similar to those described above for LIGO's optics. Additionally, a 1050\,nm bandpass optical filter and SLED on/off image subtraction are used to reduce contamination by blackbody radiation from the heaters and oven in the analysis. The images are then processed to provide scattered light scalar data (Bidirectional Reflectance Distribution Function~\cite{Stover:2012book}) and video that contain information about the growth of point scatterers as the sample is annealed. Any suspected crystal growth that is seen will be verified by comparison with crystal growth rate theory (e.g., ~\cite{MARSEGLIA198031}) and through X-ray diffraction measurements.  

%\begin{equation}
%BRDF = \frac{P_{s}}{P_{i}\cos{\theta_s}\Omega}
%\label{eq:brdf}
%\end{equation}

\section{Initial Results and Status}
\begin{figure}[ht]
\begin{minipage}[t]{0.55\textwidth}
	\includegraphics[width=1.0\textwidth]{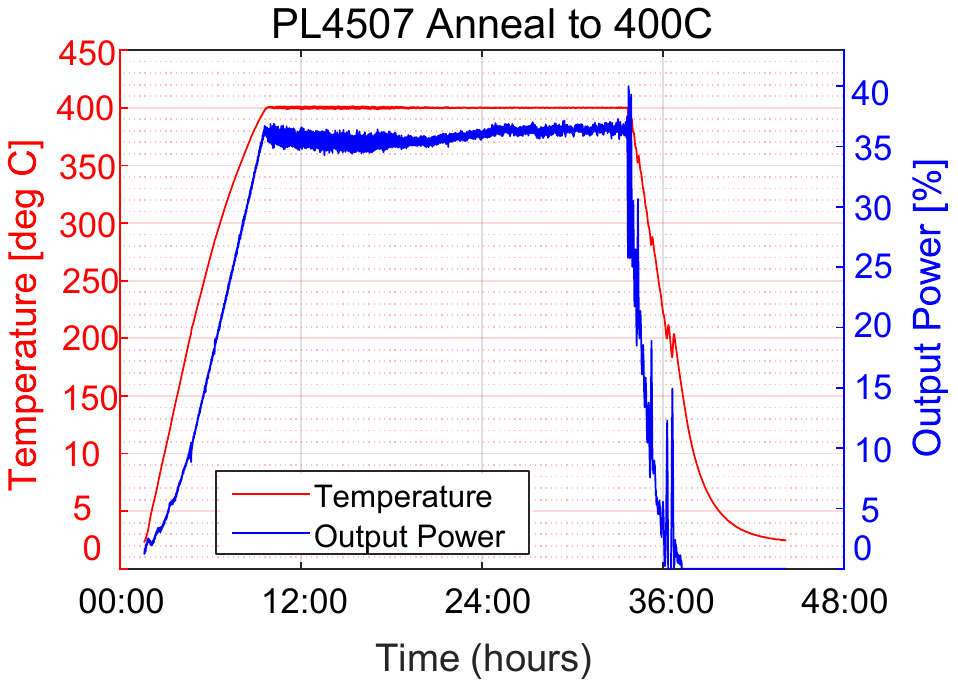}
\end{minipage}
\begin{minipage}[t]{0.45\textwidth}
	\includegraphics[width=1.0\textwidth]{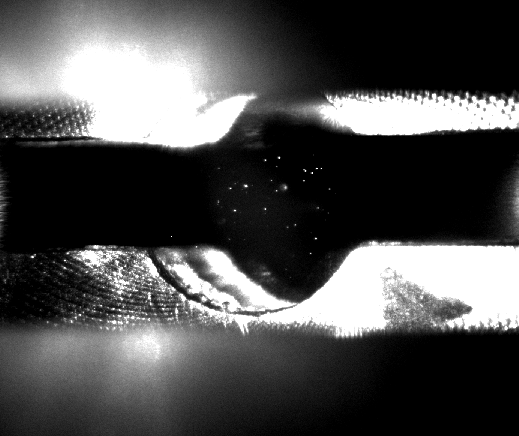}
\end{minipage}
\caption{Left: Annealing temperature profile and heater output power versus time for a representative initial experiment. Right: Image of a sample optic during an experimental run. The blurry foreground is the ceramic sheath while the textured metal is the oven's faceplate. At center, some point-like scatter is visible where the optic is illuminated. }\label{fig:results}
\end{figure}

The apparatus has been commissioned and tested with ramp and soak annealing up to 400$^{\circ}$C. The first samples tested were 1-inch fused silica substrates coated with a single quarter-wave layer of Ti:Ta$_2$O$_5$ and not previously annealed. Figure~\ref{fig:results} shows a typical annealing temperature ramp profile and a typical image of the scattering from a sample optic. In these experiments, no increase in scatter was seen over the course of annealing (as expected). 

These initial runs presented two primary challenges, especially at at higher temperatures. The first was contamination of the images by blackbody radiation. This was addressed by installing an optical bandpass filter at the entrance to the CCD camera and by illumination on/off image subraction. The second challenge was accumulation of contaminants on the optics and the viewports due to outgassing of the heated components. This was addressed with improved cleaning procedures including conducting baking runs with a dummy optic followed by chamber and viewport wipe-downs before installing the sample optic. With these improvements in place, work is now underway to reach the target 600$^{\circ}$C temperature. 

\section{Conclusion}
We have developed an experiment to measure how the optical scatter of amorphous thin-film (Ti:)Ta$_2$O$_5$ coatings changes with annealing temperature. 
% Additionally, X-ray diffraction measurements will be performed before and after annealing to assess crystallization. 
Initial results for annealing profiles that reach 400$^{\circ}$C temperatures show no scatter increase, but showed the viability of the setup and presented challenges with blackbody radiation and cleanliness that have been met. Results to higher temperatures should answer the question of whether annealing is responsible for crystal growth in LIGO's optical coatings. This information will inform the preparation of future optical coatings for low-scatter applications, especially LIGO.  

%Possibly add laser vs  SLED results.

%\section{Acknowledgements}
%The authors thank the LIGO Scientific Collaboration for discussions on this work and for review of this manuscript and acknowledge funding support NSF (PHY-1255650, HRD-1302873, AST-1559694) and  Dan O. Black and family.

%\section{References}
\bibliographystyle{osajnl} %use OSA bibliography style
\bibliography{references} % Bibliography

\end{document}